\documentclass[aps,twocolumn,prl,showpacs,superscriptaddress,floatfix]{revtex4}
\usepackage{amssymb,amsmath,amsfonts}
\usepackage{epsfig,graphicx}

\newcommand{\lv}       {\ensuremath{\left|}}
\newcommand{\rv}       {\ensuremath{\right|}}
\newcommand{\lk}       {\ensuremath{\left\{}}
\newcommand{\rk}       {\ensuremath{\right\}}}
\newcommand{\lr}       {\ensuremath{\left\langle}}
\newcommand{\rr}       {\ensuremath{\right\rangle}}
\newcommand{\be}       {\begin{equation}}
\newcommand{\ee}       {\end{equation}}

\newcommand{\xsim}[1]  {\overset{#1 \gg 1}{\sim}}
\newcommand{\xnsim}[1] {\overset{#1 \gg 1}{\nsim}}

\newcommand{\ket}[1]   {\ensuremath{\lv #1 \rr}}
\newcommand{\bra}[1]   {\ensuremath{\lr #1 \rv}}

\begin{document}
\title{Excited-state phase transition leading to symmetry-breaking
  steady states in the Dicke model}

\author{Ricardo Puebla}
\affiliation{Grupo de F\'isica Nuclear, Departamento de F\'{\i}sica At\'omica, Molecular y Nuclear, Universidad Complutense de Madrid, Av. Complutense s/n, 28040 Madrid}
\author{Armando Rela\~no}
\affiliation{Departamento de F\'{\i}sica Aplicada I and GISC, Universidad Complutense de Madrid, Av. Complutense s/n, 28040 Madrid}
 \author{Joaqu\'in Retamosa}
\affiliation{Grupo de F\'isica Nuclear, Departamento de F\'{\i}sica At\'omica, Molecular y Nuclear, Universidad Complutense de Madrid, Av. Complutense s/n, 28040 Madrid}

\begin{abstract}
  We study the phase diagram of the Dicke model in terms of the
  excitation energy and the radiation-matter coupling constant
  $\lambda$. Below a certain critical value $\lambda_c$ of the
  coupling constant $\lambda$ all the energy levels have a well-defined
  parity. For $\lambda > \lambda_c$ the energy spectrum exhibits two
  different phases separated by a critical energy $E_c$ that proves to
  be independent of $\lambda$. In the upper phase, the energy levels
  have also a well defined parity, but below $E_c$ the energy levels
  are doubly degenerated. We show that the long-time behavior of
  appropriate parity-breaking observables distinguishes between the
  different phases of the energy spectrum of the Dicke model. Steady
  states reached from symmetry-breaking initial conditions restore
  the symmetry only if their expected energies are above the
  critical. This fact makes it possible to experimentally explore the
  complete phase diagram of the excitation spectrum of the Dicke
  model.
\end{abstract}

\maketitle

{\em Introduction.-} The rapid development of experimental techniques
controlling ultra-cold atoms has given rise to a great breakthrough in
the physics of quantum many-body systems. A logical outcome has been
the increase of the interest in certain phenomena, such as
non-equilibrium dynamics and quantum phase transitions (QPTs). Also,
it has entailed the revival of well-known physical models as that
formulated by R. Dicke, which describes the interaction between an
ensemble of two-level atoms with a single electromagnetic field mode,
as a function of the radiation-matter coupling \cite{Dicke:54}. Its
most representative features are a second order QPT, which leads the
system from a normal to a superradiant phase, characterized by a
macroscopic population of the upper atomic level \cite{Hepp:73}, the
emergence of quantum chaos and the spontaneous symmetry breaking
\cite{Emary:03,Baumann:11}. Although this model has been extensively
studied from many points of view, there still exists a heated
controversy about its significance in real physical systems. A no-go
theorem was formulated in the seventies, stating that the superradiant
phase transition cannot occur in a general system of atoms or
molecules interacting with a finite number of radiation modes in the
dipole approximation \cite{Rzazewsky:75}. In addition, it is not clear
if this theorem also forbids the superradiant transition in other
realizations of the Dicke model, like in circuit QED
\cite{Nataf:10}. On the contrary, this transition
has been experimentally observed with a superfluid gas in an optical
cavity, giving rise to a self-organized phase \cite{Baumann:10}.  All
these facts have turned the Dicke model into a multidisciplinary hot
topic, involving different branches of physics. As a consequence,
there exists an intense theoretical research; a few representative
examples concern non-equilibrium QPTs \cite{Bastidas:12}, thermal
phase transitions in the ultrastrong-coupling limit \cite{Alcalde:12},
or equilibration and macroscopic quantum fluctuations
\cite{Altland:12}.

In this Letter we explore the phase diagram of the Dicke model as a
function of two control parameters: the radiation-matter coupling
constant $\lambda$ and the energy $E$ of its eigenstates. We show that
the energy spectrum can be divided into three different sectors or
phases separated by certain critical values $\lambda_c$ and $E_c$. For
$\lambda < \lambda_c$ we find that parity is a well defined quantum
number at any excitation energy. The situation is rather different if
$\lambda > \lambda_c$. Below a certain critical energy $E_c$ all the
energy levels of the system are doubly degenerated, and, as a
consequence, the parity symmetry of each level can be broken. Above
the critical energy $E_c$ there are no such degeneracies and parity is
again a good quantum number. To some extent we can say that beyond
$\lambda_c$ the excited energy levels up to energy $E_c$ inherit the
properties of the superradiant phase, characteristic of the ground
state.  We also show that this phase diagram entails measurable
effects in the long-time dynamics of certain observables.  Indeed, if
one prepares the system in a symmetry-breaking ground state of the
superradiant phase and performs a quench to a non-equilibrium state,
then the symmetry of the final steady state remains broken if its
energy is below $E_c$, whilst it is restored in the opposite case. As
a consequence, parity non-conserving observables relax to steady
values different from zero only if the energy of the non-equilibrium
state is below the critical. This fact constitutes an unheralded
characteristic of the Dicke model, that can be accessible to
experiments and shed some light over the current controversy about the
relevance of the critical behavior of this model in real physical
systems.

{\em The Dicke model.-} The Dicke Hamiltonian can be written as follows
\begin{eqnarray}
  \label{eq:H}
  H = \omega_0 J_z + \omega a^{\dagger}a + \frac{2\lambda}{\sqrt{N}}\left(a^{\dagger}+a\right)J_x,
\label{eq:Dicke}
\end{eqnarray}
where $a^{\dagger}$ and $a$ are the usual creation and annihilation
operators of photons, $\vec{J} = \left(J_x,J_y,J_z \right)$ is the
angular momentum, with a pseudo-spin length $J=N/2$, and $N$ is the
number of atoms. The frequency of the cavity mode is represented by
$\omega$ and the transition frequency by $\omega_0$. Finally, the
parameter $\lambda$ is the radiation-matter coupling. Throughout all
this Letter, we take $\hbar=1$, and $\omega = \omega_0 = 1$.  The
parity $\Pi=e^{i\pi \left(J+J_z+a^{\dagger}a \right)} $ is a conserved
quantity, due to the invariance of $H$ under $J_x \rightarrow -J_x$
and $a\rightarrow -a$ \cite{Emary:03}, and thus all the eigenstates
are labeled with positive or negative parities.  The system
undergoes a second-order QPT at $\lambda_c = \sqrt{\omega
  \omega_0}/2$, which separates the so called normal phase ($\lambda <
\lambda_{c}$) from the superradiant phase ($\lambda > \lambda_{c}$)
\cite{Hepp:73}. In the latter the ground state becomes doubly
degenerated and parity can be spontaneously broken ---because of the
fluctuations, the system can evolve into one particular ground state
without a well-defined parity \cite{Baumann:11}.

It is important to note that the system has a finite number of atoms
but infinite photons, reason why it is mandatory to set in numerical
calculations a {\em cutoff} in the photon Hilbert space. The
convergence of our results is tested, checking their stability against
small increases of this {\em cutoff}.

\begin{figure}[h]
\centering
\includegraphics[width=0.45\textwidth,height=0.5\textwidth,angle=-90]{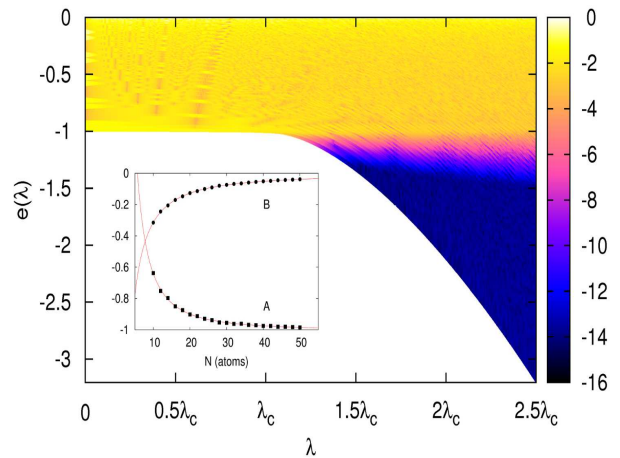}
\caption{(Color online). Intensity plot showing the decimal logarithm
  of the relative difference $\Delta E(E,\lambda,N)$ in terms of $E/J$
  and $\lambda$ for a system with $N=40$ atoms. Lighter regions
  correspond to the conserving-parity region, while darker regions
  represents the spontaneously broken-parity phase. The inset shows
  the finite-size scaling of parameters $A_N$ and $B_N$ in
  Eq. (\ref{eq:e_lineal}).}
\label{fig:Fig1}
\end{figure}

{\em Phase diagram.-} As previously commented, two different phases,
separated by $\lambda_c$, are found in Dicke model at zero
temperature. The normal phase, where parity is a well-defined quantum
number, and the superradiant phase characterized by a degenerated
ground state and a spontaneous parity-symmetry breaking. A convenient
method to see if this phenomenon is also present in excited states is
to analyze the difference

\begin{equation}
\label{eq:ae}
\Delta E_{i}(\lambda,N) = \lv \frac{E_{i}^{\Pi=+1}(\lambda,N)- E_{i}^{\Pi=-1}(\lambda,N)}{E_{i}^{\Pi=+1}(\lambda,N)}\rv 
\end{equation}
between the $i$-th excited states of both parity sectors $\Pi=\pm 1$.
If $\Delta E_{i}$ is different from zero, the corresponding
eigenstates have well-defined parity; if it is zero, they are
degenerated and one can perform a rotation that mix both parity
values. Results for $N=40$ atoms are shown in Fig. \ref{fig:Fig1}. For
$\lambda > \lambda_c$ there exists an abrupt change from $\Delta E_i
\approx 0$ to $\Delta E_i > 0$ at a certain critical energy
$E_c(\lambda,N)$. A quantitative estimate of this energy can be
obtained as the first eigenvalue $E_i$ for which $\Delta E_i >
k_{\text{err}}$, where $k_{\text{err}}$ is a given error bound. For
all the results shown below we have set $k_{\text{err}} = 10^{-6}$;
similar ones are obtained with different bounds. Since the actual
phase transition occurs in the thermodynamical limit, it is mandatory
to infer how this critical line evolves as $N \rightarrow \infty$. To do so, we
assume that for each value of $N$ the critical line obeys the linear
law

\begin{equation}
  \label{eq:e_lineal}
  \frac{E_c(\lambda,N)}{J} = A_N+B_N\lambda,  \quad \lambda>\lambda_c .
\end{equation}
where the coefficients $A_N$ and $B_N$ are numerically determined by
means of a least-squares fit. The inset of Fig. \ref{fig:Fig1} displays
the dependence of these coefficients on $N$. Solid circles represent
the numerical points corresponding to $B_N$; solid squares, the
corresponding to $A_N$, and solid lines the fits to power laws
$N^{-\alpha}$. It is clearly seen that $A_N \rightarrow -1$ and $B_N
\rightarrow 0$ as $N \rightarrow \infty$, and hence we conclude that
$E_{c}(\lambda) = E_c(\lambda,\infty) = -J$. It is worth mentioning
that this value coincides with that recently obtained in the study of
the connection of an excited-state quantum phase transition (ESQPT)
with the development of quantum chaos and the critical decay of the
survival probability \cite{Perez-Fernandez:11}.

\begin{figure}[h]
\centering
\includegraphics[width=0.45\textwidth,height=0.5\textwidth,angle=-90]{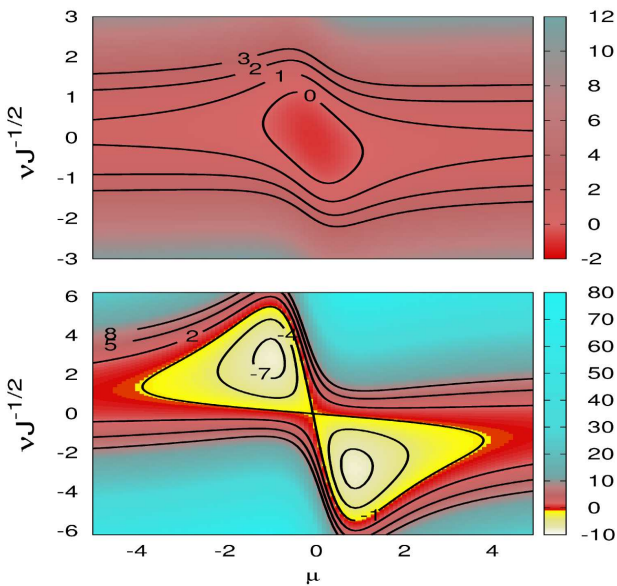}
\caption{(Color online). Contour plot of the energy surface $E_{\text{var}}
  (\mu,\nu,\lambda)/J$ for two different values of $\lambda$, one
  above and the other below the critical coupling $\lambda_c$. Upper
  panel, $\lambda=0.25$, and lower panel $\lambda=2.0$. Solid lines
  represent level curves.}
\label{fig:ener}
\end{figure}

We can also use the mean-field approximation, that gives the exact
ground-state energy in the thermodynamic limit, to estimate
$E_c(\lambda)$. Let us introduce for $\mu, \nu \in \mathbb{R}$ the
coherent ansatz $\ket{\mu,\nu} = \ket{\mu} \otimes \ket {\nu}$, where
\begin{equation}
  \label{eq:coherent}
  \begin{split}
  \ket{\mu}  & =\left(1+\mu^2\right)^{-J}e^{\mu J_+}\ket{J,-J},\\
  \ket{\nu}  & =e^{\nu^2/2}e^{\nu a^{\dagger}}\ket{0},
  \end{split}
\end{equation}
correspond to the atomic and the photonic parts of the state,
respectively.  The resulting energy surface is

\begin{equation}
  \label{eq:evar}
  \begin{split}
   E_{var}(\mu,\nu,\lambda) &= \bra{\nu,\mu} H \ket{\nu,\mu} = \omega_0 J \left( \frac{\mu^2-1}{\mu^2+1} \right) \\ 
                           &+\omega \nu^2 + \lambda \sqrt{2J}\left( \frac{4\mu \nu }{\mu^2 +1}\right).
  \end{split}
\end{equation}
It is plotted in Fig. \ref{fig:ener}; the upper panel shows the case
with $\lambda=0.25$ (below $\lambda_c$), and the lower panel, the case
with $\lambda=2.0$ (above $\lambda_c$); in both panels a number of
level curves are drawn with solid lines . The geometry of this surface
reveals that the level curve $E=-J$ plays an especial role. For
$\lambda < \lambda_c$, it reduces to a single point at
$(\mu,\nu)=(0,0)$, which is the absolute minimum of the energy
surface; for $\lambda > \lambda_c$, it changes abruptly to a
non-analytic level curve containing a saddle point. Moreover, the
shape of the energy surface is quite different depending on whether
$E$ is below or above $E = -J$. In the former case, the energy surface
exhibits two symmetric wells below $E=-J$, so that level curves are
disjoint. On the contrary, for $E > -J$ there is a single well with
connected level curves for any value of $\lambda$. This behavior
supports that $E_c(\lambda) = -J$, as our previous numerical
estimations for finite $N$.

{\em Dynamical symmetry breaking.-} Baumann and co-workers
\cite{Baumann:11} explored in real time the spontaneous parity
breaking of the ground state at the superradiant phase transition, by
measuring the behavior of $\left< J_x \right>$ as the coupling
constant $\lambda$ increases in time and crosses the critical
point. Here, we follow an analogous procedure to study the different
phases of the excitation spectrum when $\lambda > \lambda_c$. We study
the non-equilibrium dynamics and the relaxation to a steady state of
certain physical observables, like $J_x$ and $\hat{q} \equiv \left(
  a^{\dagger} + a \right) / \sqrt{2}$. They are physically measurable
operators \cite{Baumann:11, Deleglise:08}, which change the parity of
the state on which they operate. Thus, they give rise to qualitatively
different steady expectation values, depending on whether the energy
of the non-equilibrium state is above or below $E_c$. Although we only
report results for $J_x$, the behavior of $\hat{q}$ is completely
similar.

\begin{figure}[h]
\centering
\includegraphics[width=0.4\textwidth,height=0.45\textwidth,angle=-90]{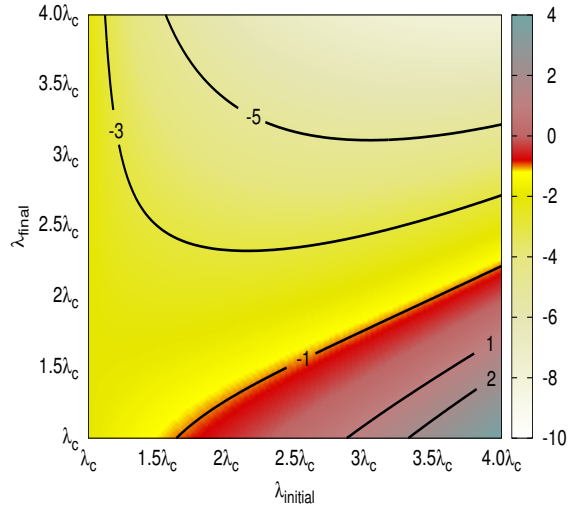}
\caption{(Color online). Contour plot of the energy surface
  $E(\lambda_i,\lambda_f)/J$ as function of $\lambda_i$ and
  $\lambda_f$. The critical energy line is placed at $E/J=-1$. The
  darker region corresponds to energies greater that the criitcal one,
  while the lighter zone corresponds to lower energies. Solid line
  represent level curves.}
\label{fig:e_lilfpic}
\end{figure}

Let us take as our initial condition a symmetry-breaking ground state
$\ket{\Psi(0)} = \ket{\mu_i,\nu_i}$, where $(\mu_i,\nu_i)$ is the
minimum of the energy surface corresponding to a coupling constant
$\lambda_{i}$ inside the superradiant phase. Then, we perform a {\em
  diabatic} change of $\lambda$, i.e., a quench
$\lambda_{i}\rightarrow \lambda_{f}$, so that the ground state
$\ket{\mu_i,\nu_i}$ becomes a non-equilibrium state of $H
\left(\lambda_f \right)$. Its energy $E(\lambda_i,\lambda_f) =
\bra{\mu_i,\nu_i} H \left(\lambda_f \right) \ket{\mu_i,\nu_i}$ can be
written as
\begin{equation}
  \label{eq:e_lilf}
  \begin{split}
    E(\lambda_i,\lambda_f)=-\omega_0J \left( \frac{\lambda_c ^2}{\lambda_i^2}\right) &+ {2J}\frac{\lambda_i ^4 -\lambda_c^4}{\omega\lambda_i^2} \\
                                                                                    &-4J\frac{\lambda_f}{\lambda_i} \left(\frac{\lambda_i^4-\lambda_c^4}{\omega\lambda_i^2} \right).
  \end{split}
\end{equation}
The contours of $E(\lambda_i,\lambda_f)$ are shown in
Fig. \ref{fig:e_lilfpic}. It is
clearly seen that choosing $\lambda_i$ and $\lambda_f$ properly, one
can explore the different phases of the excited spectrum. In
particular, from any initial initial condition satisfying that
$\lambda_i \gtrsim 1.5 \lambda_c$, both phases can be reached by just quenching
the system to different final coupling parameters $\lambda_f$.

After performing the quench, we study the time evolution of
$\lr J_x(t) \rr = \bra{\Psi(t)} J_x \ket{\Psi(t)}$, where
$\ket{\Psi(t)}=e^{-i H(\lambda_f)t}\ket{\Psi(0)}$. If one expands the
initial state in the eigenstate basis of $H \left(\lambda_f \right)$,
denoted here as $\lk \ket{E_i} \rk$, the expectation value of $J_x$
reads
\begin{equation}
  \label{eq:jx_t}
  \lr J_x (t)\rr = \sum_{i,j} C_j^{*} C_i e^{-i\left(E_i-E_j \right)t }\bra{E_j} J_x \ket{E_i},
\end{equation}
being $C_i = \lr E_i | \Psi(0) \rr$.

\begin{figure}[h]
\centering
\includegraphics[width=0.4\textwidth,height=0.45\textwidth,angle=-90]{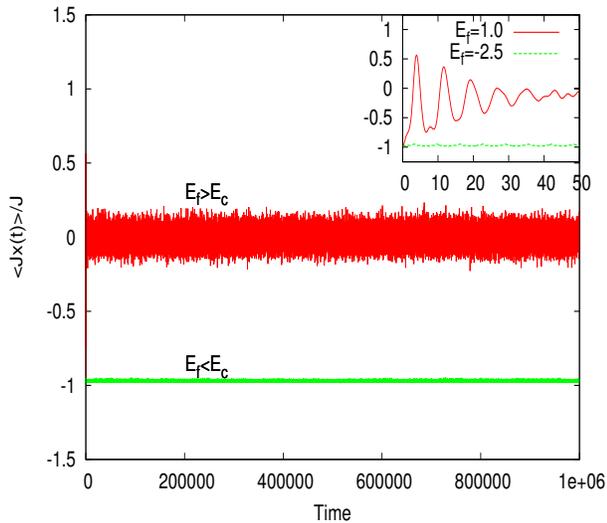}
\caption{(Color online). Time evolution of the expectation value of
  $J_x$, for two different quenches and $N=20$. Green line corresponds
  to a time evolution with $E<E_c$, and the red one to a evolution
  with $E>E_c$. The inset displays the evolution at shorter times. }
\label{fig:time_jx}
\end{figure}

Fig. \ref{fig:time_jx} displays, for a system with $N=20$ atoms, the
expected values $\lr J_x (t) \rr$ after applying two different
quenches. In the first quench $\lambda_i=1.41,\; \lambda_f=1.13$, and
the energy $E(\lambda_i,\lambda_f)/J = -2.5$ is well inside the
parity-breaking phase. It is clear seen that $\lr J_x (t) \rr$ relaxes
very quickly to a non-zero value. The same result is obtained in all
the cases where the energy of the non-equilibrium state is
$E(\lambda_i,\lambda_f) < -J$. Although in this particular case $\lr
J_x (t) \rr<0$, both positive and negative expectation values can be
obtained. For every $\lambda_i > \lambda_c$ there are two degenerate
ground states $\ket{\pm\mu_i,\pm\nu_i}$, characterized by values of
$\mu$ and $\nu$ with opposite signs, that lead to different signs of
$\lr J_x(t) \rr$.  The starting point of the second quench is also
$\lambda_i=1.41$, but the final coupling constant is reduced to
$\lambda_f = 0.51$. The energy of the non-equilibrium state,
$E(\lambda_i,\lambda_f)/J = 1.0$, is now well above the critical
energy. In this case, and in all that $E> E_c$, we obtain $\lr J_x (t)
\rr = 0$ in the steady state.

The physical explanation of this result is the following.  For
long-enough time evolutions, almost any initial condition relaxes to a
certain steady state, around which it fluctuates \cite{Reimann:12}.
Moreover, when the energy eigenvalues are not degenerated, the
expectation values of any observable $O$ in the steady state are
described by the diagonal approximation $\lr O(t) \rr \xsim{t} \lr O
\rr_D =\sum_i \left|C_i \right|^2 \bra{E_i} O \ket{E_i}$. On the
contrary, if the energy eigenvalues are degenerated, the diagonal
approximation does not hold, and thus it is possible that $\lr O (t)
\rr \xnsim{t} \lr O \rr_D$.  These are precisely the cases that we
have found in our model. In the preserving-parity phase ($E>E_c$) one
can apply the diagonal approximation because the energy levels are not
degenerated. As parity is a good quantum number in this case,
$\bra{E_i} J_x \ket{E_i} = 0$ for every energy level above
$E_c$. Therefore, whenever $E> E_c$ we find that $\lr J_x (t) \rr
\xsim{t} \lr J_x \rr_D =\sum_i \left|C_i \right|^2 \bra{E_i} J_x
\ket{E_i} = 0$.  On the contrary, in the broken-parity phase, the
energy eigenvalues are doubly degenerated with opposite parities, so
that the diagonal approximation is not
valid. Thus, one can find expectation values $\lr J_x (t) \rr
\xnsim{t} 0$ in this phase.

Consequently, the steady expected value of $J_x$ provides a neat
signature of the two phases of the excited spectrum whenever $\lambda
> \lambda_c$. In fact, it acts like an order parameter of the ESPQT,
as it is equal to zero if $E>E_c$, and different from zero if
$E<E_c$. Therefore, it suffices to follow the long-time dynamics of a
parity-changing operator to infer whether the energy of the initial
state is above or below the critical energy. Furthermore, as this is
already true for small values of $N$, finite precursors of this phase
transition could be clearly observed in experiments. In particular,
the setup used in \cite{Baumann:11} is a good candidate for covering
this aim.

{\em Conclusions.-} We have studied the phase diagram of the Dicke
model in terms of the coupling constant $\lambda$ and the energy $E$.
Using numerical calculations and the mean-field approximation, we have
found different phases in the excitation spectrum, separated by
certain critical values $E_c$ and $\lambda_c$ , where the latter also
defines the critical point of the superradiant transition of the
ground state.  For $\lambda < \lambda_c$ we find a single phase where
parity is a well defined quantum number.  On the contrary, for
$\lambda > \lambda_c$ there exists a critical energy $E_c = -J$, such
that below $E_c$ all the energy levels of the system are doubly
degenerated and composed of states with opposite parities. As a
consequence, fluctuations can entail a spontaneous parity breaking
---the system can evolve into a state without a definite value of the
parity. In some sense the excited energy levels up to energy $E_c$
inherit the properties of the ground state in the superradiant phase.
This fact leads to measurable dynamical consequences. Starting from a
symmetry-breaking ground state in the superradiant phase and
performing a quench to a non-equilibrium state, the relaxed expected
value of certain observables, like $J_x$ or $q$, is different from
zero only if the energy of the non-equilibrium state is below $E_c$.
This constitutes a new feature of the Dicke model, which could be
observed in experiments similar to that of Ref. \cite{Baumann:11}. We
think that the results contained in this Letter might shed some light
about the significance of the critical behavior of the Dicke model in
real physical systems.

{\em Aknowledgments.-} The authors thank Borja Peropadre for his
valuable comments. This work is supported in part by Spanish
Government grants for the research projects FIS2009-11621-C02-01,
FIS2009-07277, CSPD-2007-00042-Ingenio2010, and by the Universidad
Complutense de Madrid grant UCM-910059.

\end{document}